\documentclass[aps,showpacs,preprintnumbers,amsmath,amssymb,nofootinbibt,preprint]{revtex4}
\usepackage{graphicx}

\begin{document}


\title{Thin accretion disks in $f(R)$ modified gravity models}

\author{C. S. J. Pun$^1$}
\email{jcspun@hkucc.hku.hk}
\author{Z. Kov\'{a}cs$^{2,3}$}
\email{zkovacs@mpifr-bonn.mpg.de}
\author{T. Harko$^1$}
\email{harko@hkucc.hku.hk} \affiliation{$^{1}$Department of
Physics and Center for Theoretical and Computational Physics, The
University of Hong Kong, Pok Fu Lam Road, Hong Kong}
\affiliation{$^{2}$Max-Planck-Institute f\"{u}r Radioastronomie,
Auf dem H\"{u}gel 69, 53121 Bonn, Germany}
\affiliation{$^{3}$Department of Experimental Physics, University of Szeged, D\'{o}m T%
\'{e}r 9, Szeged 6720, Hungary}
\date{\today}

\begin{abstract}
We consider the basic physical properties of matter forming a thin
accretion disc in the static and spherically symmetric space-time
metric of the vacuum $f(R)$ modified gravity models. The
Lagrangian of the generalized gravity theory is also obtained in a
parametric form, and the conditions of the viability of the model
are discussed. The exact Schwarzschild type solution of the
gravitational field equations in the $f(R)$ gravity contains a
linearly increasing term, as well as a logarithmic correction, as
compared to the standard Schwarzschild solution of general
relativity, and it depends on four arbitrary integration
constants. The energy flux and the emission spectrum from the
accretion disk around the $f(R)$ gravity black holes are obtained,
and they are compared to the general relativistic case. Particular
signatures can appear in the electromagnetic spectrum, thus
leading to the possibility of directly testing modified gravity
models by using astrophysical observations of the emission spectra
from accretion disks.
\end{abstract}

\maketitle

\section{Introduction}

Several recent astrophysical observations \cite{Ri98} have
provided the astonishing result that around $95$--$96\%$ of the
content of the Universe is in the form of dark matter $+$ dark
energy, with only about $4$--$5\%$ being represented by baryonic
matter. More intriguing, around $70\%$ of the energy-density may
be in the form of what is called ``dark energy'', and is
responsible for the acceleration of the distant type Ia supernovae
\cite {PeRa03}. Hence, today's models of astrophysics and
cosmology face two severe problems, that can be summarized as the
dark energy problem, and the dark matter problem, respectively.
Although in recent years many different suggestions have been
proposed to overcome these issues, a satisfactory answer has yet
to be obtained.

A very promising way to explain these major problems is to assume
that at large scales the Einstein gravity model of general
relativity breaks down, and a more general action describes the
gravitational field. Theoretical models in which the standard
Einstein-Hilbert action is replaced by an arbitrary function of
the Ricci scalar $R$, first proposed in \cite{Bu70}, have been
extensively investigated lately. The only restriction imposed on
the function $f$ is to be analytic, that is, it must possess a
Taylor series expansion about any point. Cosmic acceleration can
be explained by $f(R)$ gravity \cite{Carroll:2003wy}, and the
conditions of viable cosmological models have been derived in
\cite{viablemodels}. In the context of the Solar System regime,
severe weak field constraints seem to rule out most of the models
proposed so far \cite{solartests,Olmo07}, although viable models
do exist
\cite{Hu:2007nk,solartests2,Sawicki:2007tf,Amendola:2007nt}. The
possibility that the galactic dynamic of massive test particles
can be understood without the need for dark matter was also
considered in the framework of $\ f(R)$ gravity models \cite
{Cap2,Borowiec:2006qr,Mar1,Boehmer:2007kx,Bohmer:2007fh}, and
connections with MOND and the Pioneer anomaly further explored by
considering an explicit coupling of an arbitrary function of $R$
with the matter Lagrangian density
\cite{Bertolami:2007gv,Bertolami:2007vu}. For a recent review of
the $f(R)$ modified gravity models see \cite{rev}.

The study of the static spherically symmetric vacuum solutions of
the gravitational field theories is fundamental for the physical
understanding and interpretation of the model. In particular, the
vacuum solutions provide the theoretical basis for the Solar
System testing of the theories, and for the description of the
motion of the test particles around massive bodies. It was shown
that for a large class of models, including e.g.~the $f(R)=R-\mu
^{4}/R$ model, the Schwarzschild-de Sitter metric is an exact
solution of the field equations.  Solutions in the presence of a
perfect fluid were analyzed in \cite{Multamaki:2006ym}. Other
approaches in searching for exact spherically symmetric solutions
of $f(R)$ theories of gravity were studied in \cite
{Capozziello:2007wc}, respectively.

Several exact vacuum static and spherically symmetric solutions of
the gravitational field equations in $f(R)$ gravity were obtained
in \cite{Multamaki:2006zb}. The set of the modified Einstein's
field equations were reduced to a single, third order differential
equation, and it was shown how one can construct exact solutions
in different $f(R)$ models. In particular, a Schwarzschild type
solution of the field equation was constructed. This solution,
containing a term linearly increasing with the radial coordinate,
as well as a logarithmic term, depends on four new arbitrary
integration constants, and reduces to the standard general
relativistic case by imposing the zero value to one integration
constant, and by appropriately choosing the numerical values of
the other constants. However, even that this choice of constants
allows the model to pass the solar system tests, we expect that at
large distances, or in the presence of strong gravitational
fields, the geometry of the space-time in $f(R)$ modified gravity
models is different from that of standard general relativity.
Therefore, it is important to find a method that could allow one
to observationally distinguish, and test in an astrophysical
setting, the possible deviations from Einstein's theory. One such
possibility is the study of accretion disks around compact
objects.

Most of the astrophysical bodies grow substantially in mass via
accretion. Recent observations suggest that around almost all of
the active galactic nuclei (AGN's), or black hole candidates,
there exist gas clouds surrounding the central compact object,
together with an associated accretion disc, on a variety of scales
from a tenth of a parsec to a few hundred parsecs \cite{UrPa95}.
These gas clouds are assumed to form a geometrically and optically
thick torus (or warped disc), which absorbs most of the
ultraviolet radiation and the soft X-rays. The gas exists in
either the molecular or the atomic phase. The most powerful
evidence for the existence of super massive black holes comes from
the VLBI imaging of molecular $\mathrm{H_{2}O}$ masers in the
active galaxy NGC 4258 \cite {Miyo95}. This imaging, produced by
Doppler shift measurements assuming Keplerian motion of the
masering source, has allowed a quite accurate estimation of the
central mass, which has been found to be a $3.6\times
10^{7}M_{\odot }$ super massive dark object, within $0.13$
parsecs. Hence, important astrophysical information can be
obtained from the observation of the motion of the gas streams in
the gravitational field of compact objects.

The mass accretion around rotating black holes was studied in
general relativity for the first time in \cite{NoTh73}.  By using
an equatorial approximation to the stationary and axisymmetric
space-time of rotating black holes, steady-state thin disk models
were constructed, extending the theory of non-relativistic
accretion \cite{ShSu73}. In these models hydrodynamical
equilibrium is maintained by efficient cooling mechanisms via
radiation transport, and the accreting matter has a Keplerian
rotation. The radiation emitted by the disk surface was also
studied under the assumption that black body radiation would
emerge from the disk in thermodynamical equilibrium. The radiation
properties of the thin accretion disks were further analyzed  in
 \cite{PaTh74} and in \cite{Th74}, where the effects of the photon
capture by the hole on the spin evolution were presented as well.
In these works the efficiency with which black holes convert rest
mass into outgoing radiation in the accretion process was also
computed.

Later on, the emissivity properties of the accretion disks were
investigated for exotic central objects, such as non-rotating or
rotating quark, boson or fermion stars  \cite{Bom,To02,YuNaRe04}.
The radiation power per unit area, the temperature of the disk and
the spectrum of the emitted radiation were given, and compared
with the case of a Schwarzschild black hole of an equal mass.

It is the purpose of the present paper to study the thin accretion
disk models applied for black holes in $f(R)$ modified gravity
models, and carry out an analysis of the properties of the
radiation emerging from the surface of the disk.

The present paper is organized as follows. The $f(R)$ gravity generalization
of the Schwarzschild type solution of general relativity is obtained in
Section II. In Section III we review the formalism of the thin disk
accretion onto compact objects. The basic properties of matter forming a
thin accretion disc in the space-time metric of the $f(R)$ modified gravity
models are considered in Section IV. We discuss and conclude our results in
Section V. In the present paper we use a system of units so that $c=G=\hbar
=k_{B}=1$, where $k_{B}$ is Boltzmann's constant.

\section{Vacuum field equations in $f(R)$ gravity}

In the $f(R)$ gravity models the gravitational action is given by
\begin{equation}
S=\int f(R)\sqrt{-g}d^{4}x,
\end{equation}
where $f(R)$ is an arbitrary functions of the Ricci scalar $R$. Since we are
only interested in the vacuum case we do not add a matter Lagrangian to the
action.

Varying the action with respect to the metric $g_{\mu \nu }$ yields the
field equations, given by
\begin{equation}
F(R)R_{\mu \nu }-\frac{1}{2}f(R)g_{\mu \nu }-\nabla _{\mu }\nabla _{\nu
}F(R)+g_{\mu \nu }\square F(R)=0,  \label{field}
\end{equation}
where we have denoted $F(R)=df(R)/dR$. Note that the covariant derivative of
these field equations vanishes for all $f(R)$ by means of the generalized
Bianchi identities \cite{Bertolami:2007gv,Koivisto}.

In the following we will restrict our study of $f(R)$ gravity models to the
static and spherically symmetric metric given by
\begin{equation}
ds^{2}=-e^{\nu (r)}dt^{2}+e^{\lambda (r)}dr^{2}+r^{2}\left( d\theta
^{2}+\sin ^{2}\theta d\phi ^{2}\right) .  \label{metr1}
\end{equation}
For a metric of the form given by Eq.~(\ref{metr1}), the vacuum
field equations of $f(R)$ gravity can be expressed as
\cite{Multamaki:2006zb}
\begin{equation}\label{f1}
\frac{F^{\prime \prime }}{F}-\frac{1}{2}(\nu ^{\prime }+\lambda ^{\prime })%
\frac{F^{\prime }}{F}-\frac{\nu ^{\prime }+\lambda ^{\prime
}}{r}=0,
\end{equation}
\begin{equation}\label{f2}
\nu ^{\prime \prime }+\nu ^{\prime }{}^{2}-\frac{1}{2}(\nu ^{\prime
}+\lambda ^{\prime })\left(\nu ^{\prime }+\frac{2}{r}\right)-\frac{2}{r^{2}}(1-e^{\lambda })=-2%
\frac{F^{\prime \prime }}{F}+\left(\lambda ^{\prime
}+\frac{2}{r}\right)\frac{F^{\prime }}{F},
\end{equation}
\begin{equation}\label{f3}
\frac{f}{F}e^{\lambda }=-\nu ^{\prime \prime }-\frac{1}{2}(\nu
^{\prime }-\lambda ^{\prime })\nu ^{\prime }+\frac{2}{r}\lambda
^{\prime }+\left(\nu ^{\prime }+\frac{4}{r}\right)\frac{F^{\prime
}}{F},
\end{equation}
\begin{equation}\label{f4}
R=2\frac{f}{F}-3e^{-\lambda }\left[ \frac{F^{\prime \prime }}{F}+\frac{1}{2}%
\left(\nu ^{\prime }-\lambda ^{\prime
}+\frac{4}{r}\right)\frac{F^{\prime }}{F}\right] ,
\end{equation}
where the prime denotes differentiation with respect to $r$.
Introducing a new variable $\xi $ by means of the transformation
$\xi =\ln r $, the field equations Eqs.~(\ref{f1}) and~(\ref{f2})
become
\begin{equation}
\frac{F_{,\left.\right.\xi \xi }}{F}-\left[ 1+\frac{1}{2}(\nu
_{,\left.\right.\xi }+\lambda _{,\left.\right.\xi })\right]
\frac{F_{,\left.\right.\xi }}{F}-(\nu _{,\left.\right.\xi
}+\lambda _{,\left.\right.\xi })F=0\,,
\end{equation}
\begin{equation}
\nu _{,\left.\right.\xi \xi }-\nu _{,\left.\right.\xi }+\nu
_{,\left.\right.\xi }^{2}-\frac{1}{2}(\nu _{,\left.\right.\xi
}+\lambda _{,\left.\right.\xi })(\nu _{,\left.\right.\xi
}+2)+2(1-e^{\lambda })=-2\frac{F_{,\left.\right.\xi \xi
}}{F}+(\lambda _{,\left.\right.\xi }+4)\frac{F_{,\left.\right.\xi
}}{F}\,,
\end{equation}
where the comma denotes differentiation with respect to the variable $\xi $%
. As a result of introducing the new independent variable the
basic field equations Eqs.~(\ref{f1}) and~(\ref{f2}) are
independent of the coordinate $\xi $. It is useful to introduce a
formal representation of the function $F$ as $F_{,\left.\right.\xi
}/F=u$, $F_{,\left.\right.\xi \xi }/F=u_{,\left.\right.\xi
}+u^{2}$, with $u$ a new function of $\xi $. Then Eq.~(\ref{f1})
can be written as
\begin{equation}\label{h1}
u_{,\left.\right.\xi }+u^{2}-\left[ 1+\frac{1}{2}(\nu
_{,\left.\right.\xi }+\lambda _{,\left.\right.\xi })\right] u-(\nu
_{,\left.\right.\xi }+\lambda _{,\left.\right.\xi })=0.
\end{equation}
This equation is a Riccati type first order differential equation.
By using the function $u$, Eq. (\ref{f2}) becomes
\begin{equation}\label{h2}
\nu _{,\left.\right.\xi \xi }-\nu _{,\left.\right.\xi }+\nu
_{,\left.\right.\xi }^{2}-\frac{1}{2}(\nu _{,\left.\right.\xi
}+\lambda _{,\left.\right.\xi })(\nu _{,\left.\right.\xi
}+2)+2(1-e^{\lambda })=-2u_{,\left.\right.\xi }-2u^{2}+(\lambda
_{,\left.\right.\xi }+4)u.
\end{equation}

Substituting the term $u_{,\left.\right.\xi }+u^{2}$ in
Eq.~(\ref{h2}) with the use of Eq.~(\ref{h1}), we obtain
\begin{equation}\label{sol2}
2(1-e^{\lambda })+\nu _{,\left.\right.\xi \xi }-\nu
_{,\left.\right.\xi }+\nu _{,\left.\right.\xi }^{2}+(\nu
_{,\left.\right.\xi }+\lambda _{,\left.\right.\xi
})(1-\frac{1}{2}\nu _{,\left.\right.\xi })-(2-\nu
_{,\left.\right.\xi })u=0.
\end{equation}

The general solution of Eqs.~(\ref{h1}) and~(\ref{sol2}) gives the
general solution of the field equations for the static vacuum case
of the $f(R)$ gravity models. Once $\nu (\xi )$ and $\lambda (\xi
)$ are specified, one can immediately obtain $u$ and then (by
integration) $F$, as well as all the other relevant physical
quantities. If the function $F$ and the metric tensor coefficients
are known, $f$ can be obtained as a function of $R$ from
Eqs.~(\ref{f3}) and~(\ref{f4}) in a parametric form, as $f=f\left(
\xi \right) $, $R=R\left( \xi \right) $.

We now consider general solutions with $\nu _{,\left.\right.\xi
}+\lambda _{,\left.\right.\xi }=0$, which we denote as
Schwarzschild-type solutions. The constant of integration may be
set to zero by re-scaling the time coordinate, so that without a
significant loss of generality one may consider the solution $\nu
+\lambda =0 $. Thus, Eq. (\ref{h1}) reduces to
$u_{,\left.\right.\xi }+u^{2}-u=0$, which provides the solution
\begin{equation}
u=\frac{e^{\xi }}{e^{\xi }+C}=\frac{r}{r+C}\,,
\end{equation}
where $C$ is a constant of integration. Next, from the definition
of $u$, and by reverting back to the $r$ coordinate, we obtain
$F(r)=Ar+B$, with $A$ and $B$ constants of integration. Solving
Eq. (\ref{sol2}), one finds the Schwarzschild type general
solution in $f(R)$ gravity, given by
\begin{eqnarray}
e^{\nu }&=&e^{-\lambda }=1-\frac{AC_{2}}{2B^{2}}+\frac{C_{2}}{3Br}-\frac{%
A(B^{2}-AC_{2})}{B^{3}}r+\frac{1}{B^{2}}\left( A^{2}-
\frac{C_{1}}{6B^{2}}\right) r^{2}+ \nonumber\\
&&\frac{A^{2}(B^{2}-AC_{2})r^{2}}{B^{4}}\ln \left( A+\frac{B}{r}%
\right),
\end{eqnarray}
where $C_{1}$ and $C_{2}$ are arbitrary constants of integration.
An interesting difference to the vacuum solutions in general
relativity is the presence of the linear term with respect to $r$,
and of the term with the logarithmic dependence of $r$. In the
general case the Schwarzschild type exact vacuum solution in
$f(R)$ gravity depends on four arbitrary integration constants
$A$, $B$, $C_{1}$ and $C_{2}$, respectively. In order to obtain
the Schwarzschild-de Sitter solution of standard general
relativity, one sets the following values for the constants: $A=0$, $%
C_{2}=-6BM$ and $C_{1}=2B^{4}\Lambda $, respectively, where $M$ is
the mass of the central object \cite{Multamaki:2006zb}.

In the following we assume that $C_{2}=-6BM$, a condition which is
necessary to recover the standard Schwarzschild solution.
Moreover, we neglect the possible effect of a cosmological
constant by taking $C_{1}=0$. By denoting the ratio $A/B=1/\eta $,
we represent the vacuum metric in $f(R)$ gravity in the form
\begin{equation}
e^{\nu }=e^{-\lambda }=1+\frac{3M}{\eta }-\frac{2M}{r}-\left( 1+\frac{6M}{%
\eta }\right) \left( \frac{r}{\eta }\right) +\left( \frac{r}{\eta }\right)
^{2}\left\{ 1+\left( 1+\frac{6M}{\eta }\right) \ln \left[ A\left( 1+\frac{%
\eta }{r}\right) \right] \right\} \,.  \label{metreta}
\end{equation}

By using the Schwarzschild-type solution of the field equations in
the modified $f(R)$ gravity model given by Eqs. (\ref{metreta}) we
obtain for the functions $f(r)$, $R(r)$ and $dF/dR=d^{2}f/dR^{2}$
the following expressions
\begin{eqnarray}\label{cond1}
f(r)&=&\frac{1}{\eta ^{2}\left( r+\eta \right) r^{2}}\left\{
2A\eta \left[ 3\left( 6M-r\right) r^{2}+\eta r\left( 9M+r)\right)
-3\eta ^{2}\left(M-r\right)\right]\right.\nonumber\\
&&\left.-6Ar^{2}\left( 6M+\eta \right) \left( r+\eta \right) \ln \left[A\left(1+\eta r^{-1}\right)\right]%
\right\} ,
\end{eqnarray}
\begin{eqnarray}\label{cond2}
R(r)&=&\frac{1}{\eta ^{3}\left( r+\eta \right) ^{2}r^{2}}\left\{
\eta \left[ 6M\left( 2r+\eta \right) \left( 6r^{2}+6\eta r-\eta
^{2}\right) +r\left( -12r^{3}-12\eta r^{2}+7\eta ^{2}r+6\eta
^{3}\right) \right]\right.\nonumber\\
 &&\left.-12r^{2}\left( 6M+\eta \right) \left(
r+\eta \right) \ln \left[A\left(1+\eta
r^{-1}\right)\right]\right\}
\end{eqnarray}
and
\begin{equation}\label{cond3}
\frac{dF}{dR}=\frac{d^{2}f}{dR^{2}}=-\frac{Ar^{3}\left( r+\eta \right) ^{3}}{%
2r^{3}+6\eta r^{2}+6\eta ^{2}\left( r-2M\right) }
\end{equation}
respectively. Generally, the relation between $f$ and $R$ cannot
be obtained in an explicit analytical form. However, some
approximate representations are possible in the limiting cases of
small and large $r$, respectively.

In the limit of small $r$, so that $r$ is approximately equal to a few $M$, the logarithmic correction $%
\ln \left( A+B/r\right) $ becomes the dominant term, and we obtain $%
f(r)\approx -\chi _{f}^{-1}\ln B/r$, where $\chi _{f}=\eta
^{2}/6A\left( 6M+\eta \right) $, and  $R(r)\approx -\chi
_{R}\left( r+\eta \right) ^{-1}\ln B/r$, where $\chi _{R}=12\left(
6M+\eta \right) /\eta ^{3}$. By eliminating $r$ using the relation
$r=B\exp \left( -\chi _{f}f\right) $ we obtain
\begin{equation}
R=\frac{\chi _{f}\chi _{R}}{B\exp \left( -\chi _{f}f\right) +\eta
}f.
\end{equation}

If $f$ is small we reobtain the Lagrangian of the standard general
relativity, $f\sim R$. By performing a series expansion of the
exponential factor we obtain
\begin{equation}
f\left( R\right) \approx \frac{B+\eta }{\chi _{f}}\frac{R}{BR+\chi
_{R}}\approx \frac{B+\eta }{\chi _{f}\chi _{R}}R\left( 1-\frac{B}{\chi _{R}}%
R\right) .
\end{equation}

In the limit of large $r$, so that $r>>M$ and $r>>\eta $, by
keeping only
the terms in $1/r$, we obtain $f(r)\approx \chi _{f\infty }+1/\eta r$, and $%
R(r)\approx \chi _{R\infty }/r-12/\eta ^{2}$, where $\chi _{f\infty }=-6%
\left[ 6\eta +\left( 6M+\eta \right) \ln A\right] A$ and $\chi _{R\infty }=12%
\left[ 6M\eta -\eta ^{2}-\left( 6M+\eta \right) \ln A\right] /\eta
^{3}$. Eliminating $1/r$ we obtain
\begin{equation}
f(R)=\frac{1}{\eta \chi _{R\infty }}R+\Lambda ,
\end{equation}
where $\Lambda =12/\eta ^{3}\chi _{R\infty }+\chi _{f\infty }$.
Therefore at large distances the modified $f(R)$ gravity model
generates a constant term, the cosmological constant, that is
responsible for the accelerated expansion of the Universe.

Eqs.~(\ref{cond1})-(\ref{cond3}) can be used to discuss the
conditions under which the generalized Schwarzschild type
solution, as well as the corresponding $f(R)$ generalized gravity
theory, represents a viable model. The conditions under which
$f(R)$ theories can represent viable models of cosmic acceleration
have been summarized in \cite{PoSi07}. The requirement of the
existence of a stable, high curvature regime imposes the constraint $d^{2}f/dR^{2}>0$, for $%
R>>d^{2}f/dR^{2}$. From Eq. (\ref{cond3}) it follows that this
condition can be satisfied for all $r$ if $A<0$ and $\eta >0$. Moreover, the condition $%
\left( r/\eta \right) ^{3}+3\left( r/\eta \right) ^{2}+3\left(
r/\eta \right) >6M/\eta $ must also hold for all $r$, $\eta $ and
$M$. From the tight constraints of the Big Bang Nucleosynthesis
and of the Cosmic Microwave Background we require that $F=df/dR$
must be a negative, monotonically increasing function of $R$ that
asymptotically approaches to zero from below. The condition
$F=-Ar+B<0$ can be used to constrain the range of the radial
variable as $r>\eta $, where we have assumed that $B>0$.

From the requirement that the effective Newton constant
$G_{eff}=G/\left( 1+F\right) $ is not allowed to change sign we
obtain the condition $1+F>0$, at all finite $R$. This condition
gives $r<\eta +1/A$. Therefore in the present model the range of
the radial variable $r$ is restricted to $\eta <r<\eta +1/A$. In
order for the model to be viable on very large distances,
corresponding to cosmic scales, $A$ must have a very small
numerical value. Finally, $F$ must be a small quantity, a
condition which is required for the model to pass the solar and
galactic scale constraints. In terms of the scalar curvature $R$,
the conditions of viability of the $f(R)$ gravity model with
Schwarzschild like vacuum solution can be formulated, in the small
$r$ limit, as $-2B\left( B+\eta
\right) /\chi _{f}\chi _{R}^{2}<0$ (condition which follows from $%
d^{2}f/dR^{2}<0$), $\chi _{R}/2B<R<\chi _{R}/2B+\chi _{f}\chi
_{R}^{2}/2B\left( B+\eta \right) $, which follows from $df/dR<0$ and $%
1+df/dR>0$, respectively.

Since $A$ must be a very small (dimensionless) quantity, $A<<1$,
the condition of the smallness of $F$ is automatically satisfied
by also assuming a small numerical value for $B$. By assuming that
the value of $\left| F\right| $ should not exceed today the
numerical value of $10^{-6} $ at any point in the space-time
\cite{PoSi07}, the requirement that the modified gravity model is
viable on a scale of $r=1000M$, for example, where $M$ is the mass
of the central black hole, and by representing $B$ as $B=\beta M$,
where $\beta $ is a constant, gives the condition $\beta
-1000A=10^{-6}/M$. For a black hole with a mass of around three
solar masses we obtain $\beta
-1000A\approx 6.74\times 10^{-12}$. For $\beta =1$ it follows that $A$ is very close to $%
A\approx 10^{-3}$.

\section{Thin accretion disks onto black holes}

For the thin accretion disk it is assumed that its horizontal size
is negligible as compared to its vertical extension, i.e, the disk
height $H$, defined by the maximum half thickness of the disk, is
always much smaller than the characteristic radius $r$ of the
disk, $H<<r$. The thin disk is in hydrodynamical equilibrium,
where there is only negligible pressure gradient and a vertical
entropy gradient in the accreting matter. The efficient cooling
via the radiation over the disk surface prevents the disk from
cumulating the heat generated by stresses and dynamical friction.
In turn this equilibrium causes the disk to stabilize its thin
vertical size. The thin disk has an inner edge at the marginally
stable orbit of the black hole potential, and the accreting plasma
has a Keplerian motion in higher orbits.

In steady state accretion disk models, the mass accretion rate
$\dot{M}_{0}$ is assumed to be a constant that does not change
with time, and the physical quantities describing the orbiting
plasma are averaged over a characteristic time scale, e.g. $\Delta
t$, over the azimuthal angle $\Delta \phi =2\pi $ for a total
period of the orbits, and over the height $H$ \cite{ShSu73,
NoTh73,PaTh74}.

The particles moving in Keplerian orbits around the black hole
with a rotational velocity $\Omega =d\phi /dt$ have a specific
energy $\widetilde{E} $ and a specific angular momentum
$\widetilde{L\text{,}}$ which, in the steady state thin disk
model, depend only on the radii of the orbits. These particles,
orbiting with the four-velocity $u^{\mu }$, form a disk of an
averaged surface density $\Sigma $, the vertically integrated
average of the rest mass density $\rho _{0}$ of the plasma. The
accreting matter in the disk is modelled by an anisotropic fluid
source, where the density $\rho _{0}$ , the energy flow vector
$q^{\mu }$ and the stress tensor $t^{\mu \nu }$ are measured in
the averaged rest-frame (the specific heat was neglected). Then
the disc structure can be characterized by the surface density of
the disk \cite{NoTh73,PaTh74},
\begin{equation}
\Sigma(r) = \int^H_{-H}\langle\rho_0\rangle dz,
\end{equation}
with averaged rest mass density $\langle\rho_0\rangle$ over $\Delta t$ and $%
2\pi$ and the torque
\begin{equation}
W_{\phi}{}^{r} =\int^H_{-H}\langle t_{\phi}{}^{r}\rangle dz,
\end{equation}
with the averaged component $\langle t^r_{\phi} \rangle$ over
$\Delta t$ and $2\pi$. The time and orbital average of the energy
flux vector gives the radiation flux ${\mathcal F}(r)$ over the
disk surface as ${\mathcal F}(r)=\langle q^z \rangle$.

The stress-energy tensor is decomposed according to
\begin{equation}
T^{\mu \nu }=\rho _{0}u^{\mu }u^{\nu }+2u^{(\mu }q^{\nu )}+t^{\mu \nu },
\end{equation}
where $u_{\mu }q^{\mu }=0$, $u_{\mu }t^{\mu \nu }=0$. The
four-vectors of the energy and angular momentum flux are defined
by $-E^{\mu }\equiv T_{\nu }^{\mu }{}(\partial /\partial t)^{\nu
}$ and $J^{\mu }\equiv T_{\nu }^{\mu }{}(\partial /\partial \phi
)^{\nu }$, respectively. The structure equations of the thin disk
can be derived by integrating the conservation laws of the rest
mass, of the energy, and of the angular momentum of the plasma,
respectively \cite{NoTh73,PaTh74}. From the equation of
the rest mass conservation, $\nabla _{\mu }(\rho _{0}u^{\mu })=0$,
it follows that the time averaged rate of the accretion of the
rest mass is independent of the disk radius,
\begin{equation}
\dot{M_{0}}\equiv -2\pi r\Sigma u^{r}={\rm constant}.
\label{conslawofM}
\end{equation}
The conservation law $\nabla _{\mu }E^{\mu }=0$ of the energy has the
integral form
\begin{equation}
\lbrack \dot{M}_{0}\widetilde{E}-2\pi r\Omega W_{\phi }{}^{r}]_{,r}=4\pi r{\mathcal F}%
\widetilde{E}\;\;,  \label{conslawofE}
\end{equation}
which states that the energy transported by the rest mass flow, $\dot{M}_{0}%
\widetilde{E}$, and the energy transported by the dynamical
stresses in the disk, $2\pi r\Omega W_{\phi }{}^{r}$, is in
balance with the energy radiated away from the surface of the
disk, $4\pi r{\mathcal F}\widetilde{E}$. The law of the angular
momentum conservation, $\nabla _{\mu }J^{\mu }=0$, also states the
balance of these three forms of the angular momentum transport,
\begin{equation}
\lbrack \dot{M}_{0}\widetilde{L}-2\pi rW_{\phi }{}^{r}]_{,r}=4\pi r{\mathcal F}%
\widetilde{L}\;\;.  \label{conslawofL}
\end{equation}

By eliminating $W_{\phi }{}^{r}$ from Eqs. (\ref{conslawofE}) and (\ref
{conslawofL}), and applying the universal energy-angular momentum relation $%
dE=\Omega dJ$ for circular geodesic orbits in the form $\widetilde{E}%
_{,r}=\Omega \widetilde{L}_{,r}$, the flux ${\mathcal F}$ of the
radiant energy over the disk can be expressed in terms of the
specific energy, angular momentum and of the angular velocity of
the black hole \cite{NoTh73,PaTh74},
\begin{equation}
{\mathcal F}(r)=-\frac{\dot{M}_{0}}{4\pi r}\frac{\Omega
_{,r}}{(\widetilde{E}-\Omega
\widetilde{L})^{2}}\int_{r_{ms}}^{r}(\widetilde{E}-\Omega \widetilde{L})%
\widetilde{L}_{,r}dr\;\;.  \label{F}
\end{equation}

In the derivation of the above formula the ''no torque'' inner
boundary condition were prescribed, where the torque vanishes at
the inner edge of the disk. Thus, we assume that the accreting
matter at the marginally stable orbit $r_{ms}$ falls freely into
the black hole, and cannot exert considerable torque on the disk.
The latter assumption is valid only if strong magnetic fields do
not exist in the plunging region, where matter falls into the
black hole.

Another important characteristics of the mass accretion process is
the efficiency with which the central object converts rest mass
into outgoing radiation. This quantity is defined as the ratio of
the rate of the radiation of energy of photons escaping from the
disk surface to infinity, and the rate at which mass-energy is
transported to the black hole, both measured at infinity
\cite{NoTh73,PaTh74}. If all the emitted photons can escape to
infinity, the efficiency is given in terms of the specific energy
measured at the marginally stable orbit $r_{ms}$,
\begin{equation}
\epsilon =1-\widetilde{E}_{ms}.\label{epsilon}
\end{equation}

For Schwarzschild black holes the efficiency $\epsilon $ is about
$6\%$, whether the photon capture by the black hole is considered,
or not. Ignoring the capture of radiation by the hole, $\epsilon $
is found to be $42\%$ for rapidly rotating black holes, whereas
the efficiency is $40\%$ with photon capture in the Kerr potential
\cite{Th74}.

In order to compute the flux integral given by Eq.~(\ref{F}), we
determine the radial
dependence of the angular velocity $\Omega $, of the specific energy $%
\widetilde{E}$ and of the specific angular momentum
$\widetilde{L}$ of particles moving in circular orbits around
black holes in the static and spherically symmetric geometry given
by Eq.~(\ref{metr1}). The geodesic equations for particles
orbiting in the equatorial plane of the black hole can be written
as
\begin{equation}
e^{2\nu}\dot{t}^2 = \widetilde{E}^2, e^{(\nu+\lambda)}\dot{r}^2+
V_{eff}(r)  = \widetilde E^2, r^4\dot{\phi}^2 =  \widetilde{L}^2,
\end{equation}
where the dot denotes the differentiation with respect to the affine
parameter, and the effective potential is given by
\begin{equation}
V_{eff}(r)\equiv
e^{\nu}\left(1+\frac{\widetilde{L}}{r^2}^2\right)\;. \label{V2}
\end{equation}
From the conditions $V_{eff}(r)=0$ and
$V_{eff,\left.\right.r}(r)=0$ we obtain
\begin{eqnarray}
\Omega & = & \sqrt{\frac{\nu_{,r}e^{\nu}}{2r}}\;,  \label{Omega} \\
\widetilde{E} & = & \frac{e^{\nu}}{\sqrt{e^{\nu}-r^2\Omega^2}}\;,  \label{E}
\\
\widetilde{L} & = &
\frac{r^2\Omega}{\sqrt{e^{\nu}-r^2\Omega^2}}\;. \label{L}
\end{eqnarray}
The condition $V_{eff,\left.\right.rr}(r)=0$ gives the marginally
stable orbit $r_{ms}$ (or the innermost stable circular orbit),
which can be determined for any explicit expression of the
function $\nu(r)$.

It is possible to define a temperature $T(r)$ of the disk, by
using the definition of the flux, as ${\mathcal F}(r)=\sigma
T^{4}(r)$, where $\sigma $ is the Stefan-Boltzmann constant.
Considering that the disk emits as a black body, one can use the
dependence of $T$ on ${\mathcal F}$ to calculate the luminosity
$L\left( \omega \right) $ of the disk through the expression for
the black body spectral distribution \cite{To02},
\begin{equation}
L\left( \omega \right) =4\pi d^{2}I\left( \omega \right) =\frac{4}{\pi }\cos
i\omega ^{3}\int_{r_{i}}^{r_{f}}\frac{rdr}{\exp \left( \omega /T\right) -1},
\end{equation}
where $d$ is the distance to the source, $I(\omega )$ is the Planck
distribution function, $i$ is the disk inclination, and $r_{i}$ and $r_{f}$
indicate the position of the inner and outer edge of the disk, respectively.

\section{Thin disk accretion properties for $f(R)$ modified gravity black holes}

The effective potential which determines the geodesic motion of the test
particles in the equatorial plane of the metric given by Eq. (\ref{metreta})
can be written as
\begin{eqnarray}
V_{eff}(r)&=&\left( 1+\frac{\widetilde{L}^{2}}{r^{2}}%
\right) \left\{ 1+\frac{3M}{\eta }-\frac{2M}{r}-\left( 1+\frac{6M}{\eta }%
\right) \left( \frac{r}{\eta }\right)\right. + \nonumber\\
&&\left.\left( \frac{r}{\eta }\right) ^{2}+\left( \frac{r}{\eta
}\right)
^{2}\left( 1+\frac{6M}{\eta }\right) \ln \left[ A\left( 1+\frac{\eta }{r}%
\right) \right] \right\}.
\end{eqnarray}

\begin{figure}[!ht]
\centering
\includegraphics[width=.48\textwidth]{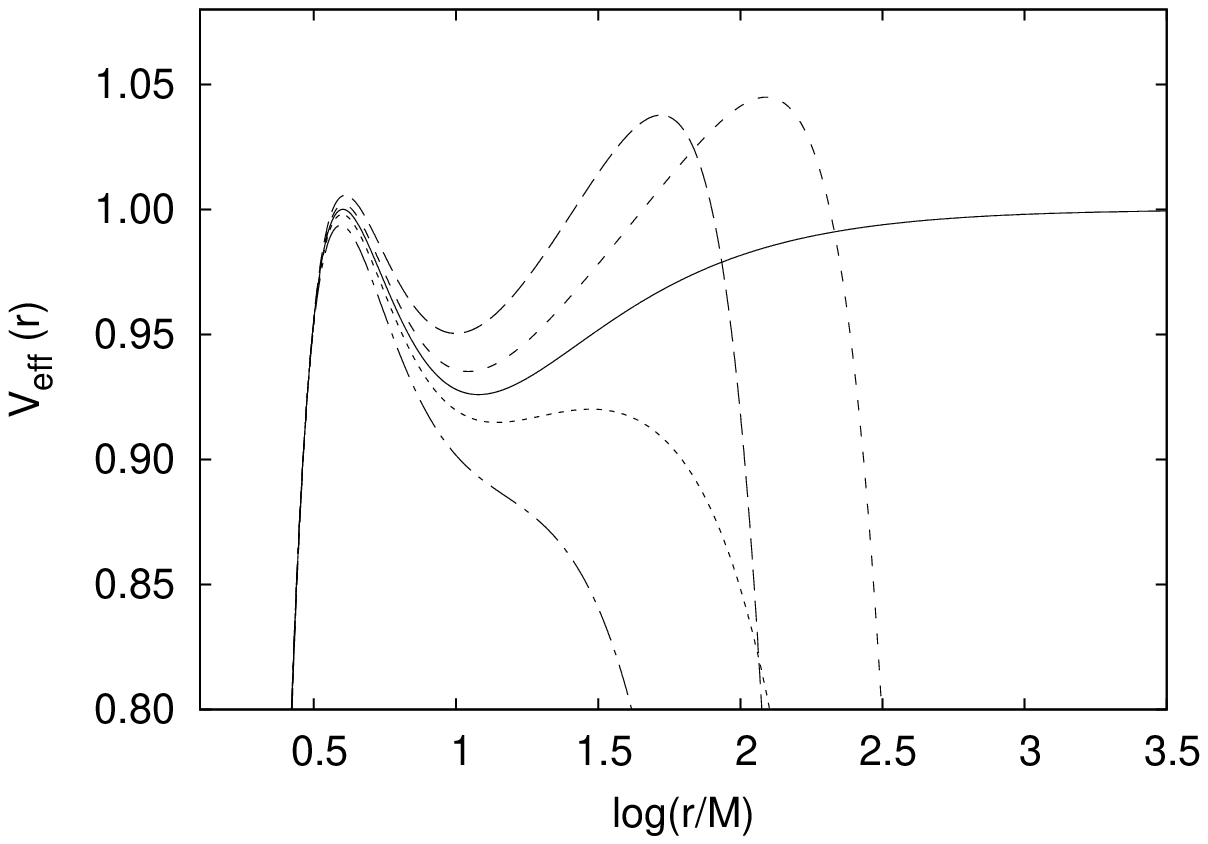}
\includegraphics[width=.48\textwidth]{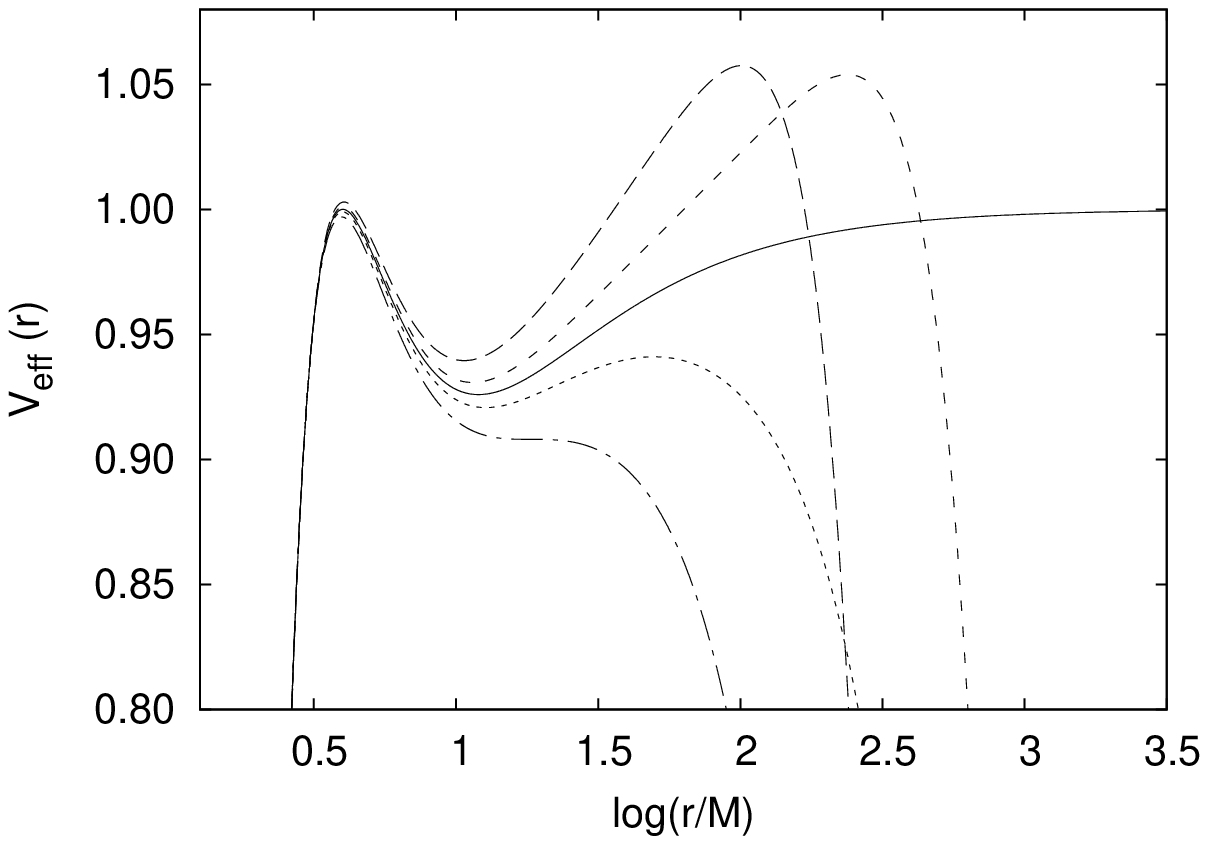}\\
\includegraphics[width=.48\textwidth]{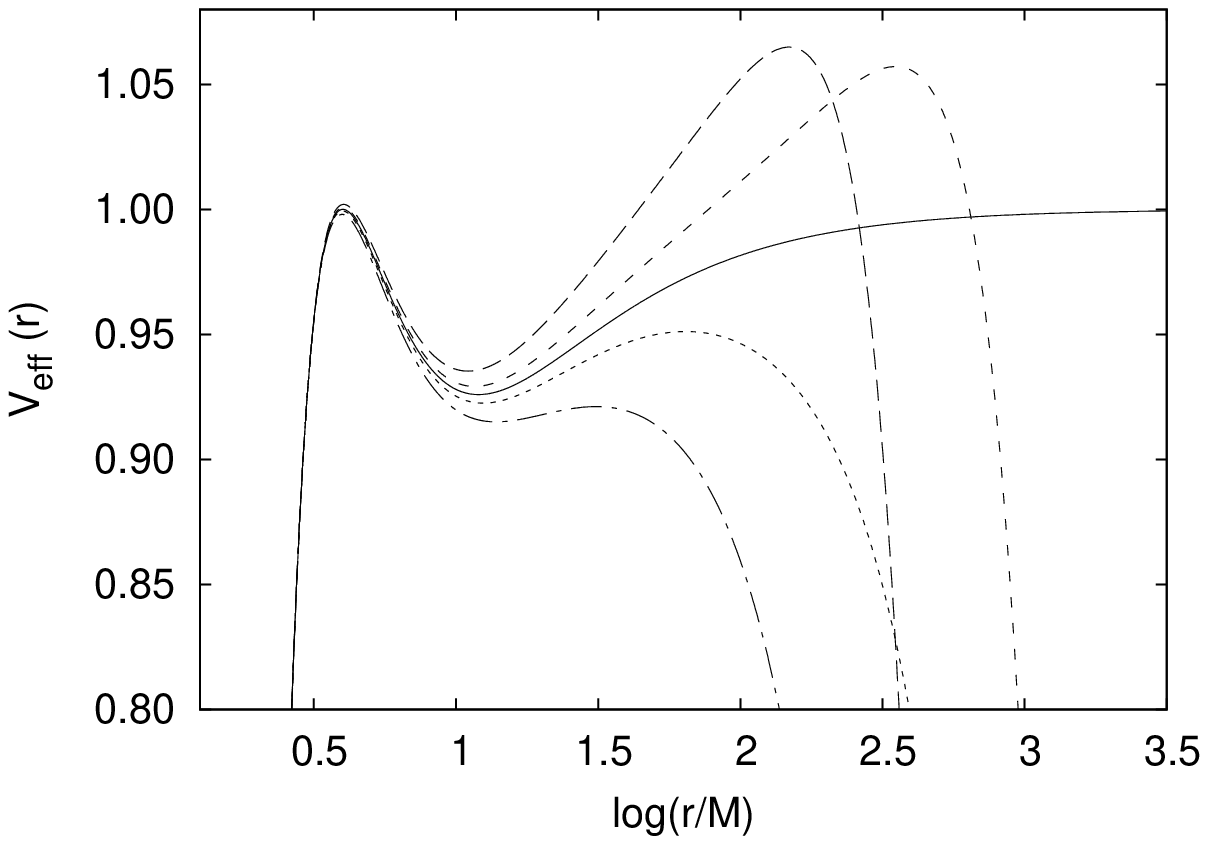}
\includegraphics[width=.48\textwidth]{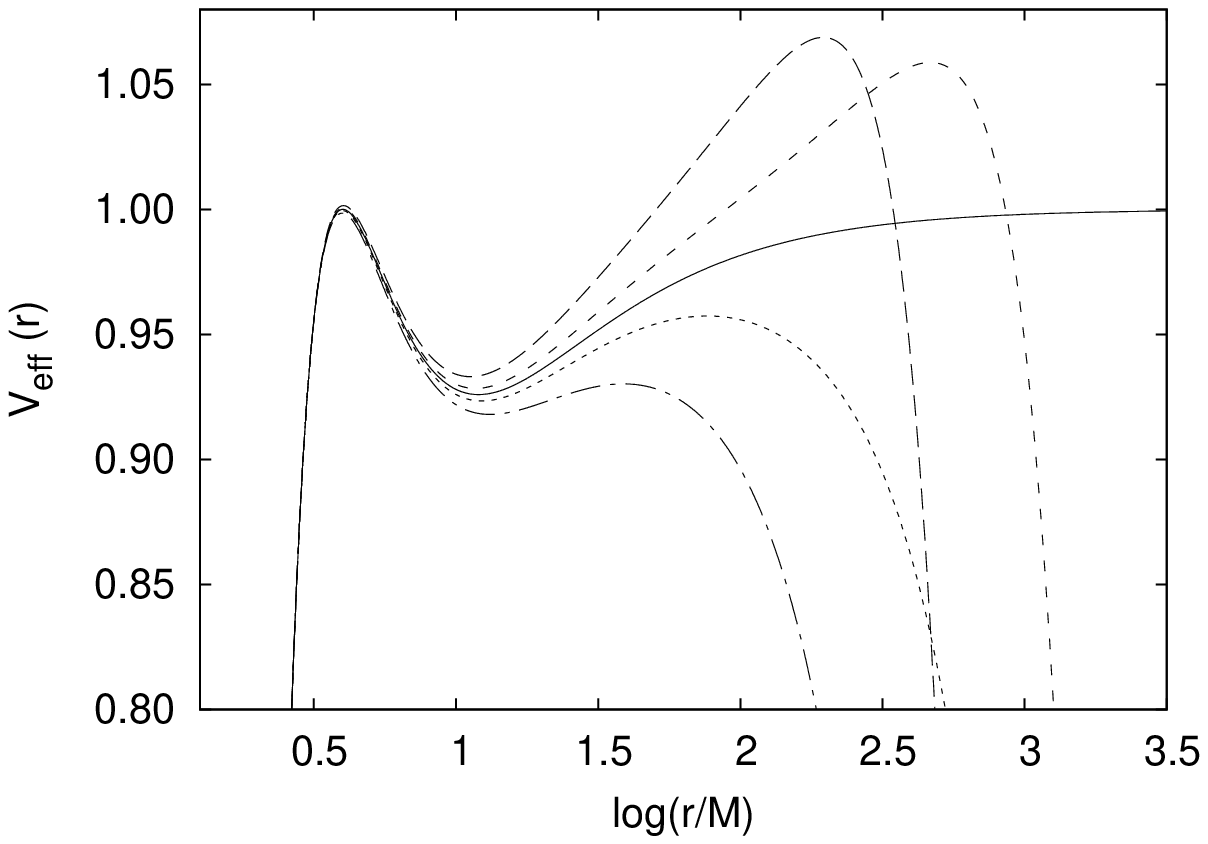}
\caption{The effective potential $V_{eff}(r)$ for a black hole of total mass $%
M=2.5\times10^6 M_{\bigodot}$ and for the specific angular momentum $%
\widetilde{L}=4M$. The metric parameter $A$ is varied in the range of $%
-3\times10^{-3}$ and $3\times10^{-3}$ for fixed $B$ in each plot.
The parameter $B$ is set to $M$ (the left hand upper plot), $2M$
(the right hand upper plot), $3M$ (the left hand lower plot) and
$4M$ (the right hand lower plot), respectively. The solid line is
the effective potential for a Schwarzschild black hole with the
 total mass $M$. The various values of $A$ are
$-3\times10^{-3}$ (long dashed line), $-10^{-3}$ (short dashed
line), $10^{-3}$ (dotted line) and $3\times10^{-3}$ (dot-dashed
line).\label{fR_V}}
\end{figure}

We plot $V_{eff}$ for different parameters $A$ (or $\eta$) in
Fig.~\ref{fR_V}. Here the parameter $A$ runs over the range of $-3\times10^{-3}$ and $%
3\times10^{-3}$ in each plot whereas $B$ is set to $M$, $2M$, $3M$
and $4M$, so that $\eta =M/A, 2M/A, 3M/A$ and $4M/A$,
respectively. The effective potential of the Schwarzschild
solution is also plotted for comparison.

The unique singular behavior of $V_{eff}(r)$ at the spatial
infinity, as opposed to the asymptotical fall-off of the
Schwarzschild potential, is the most striking feature in the upper
pairs of the plots, where all the curves representing the $f(R)$
potential for different parameters tend to $-\infty$ as
$r\rightarrow\infty$. This is only a pure manifestation of the
global properties of the spacetime for vacuum $f(R)$ gravity
models. It is also true for the cases in the lower two panels,
where the curves plotted above the Schwarzschild limit go to
$-\infty$ at the spatial infinity, after reaching a local maximum
at the marginally bound orbit. This indicates the obvious fact
that though the metric parameters considerably determine the local
properties of the gravitational field, the globally hyperbolic
geometry must be insensitive to them.

This behavior of the potential in the $f(R)$ modified gravity
models directly follows from the presence of the logarithmic
correction term in $V_{eff}$, and does not depend on the sign of
the parameters $A$ and $B$. In the case of a small negative value
of $A$, for some critical value of $r$ the argument of the
logarithm tends to zero, and the logarithmic term dominates all
the other terms in the potential, which behaves like
$V_{eff}(r)\approx\left(r/\eta \right)^2\left(1+6M/\eta
\right)\ln\left(-A+B/r\right)$ (the minus sign is due to the
smallness of the argument of the logarithmic function $-A+B/r$).
Interestingly, the same behavior appears if $A$ is a small
positive number. In this case for $B/r\rightarrow0$, the potential
behaves like $V_{eff}(r)\approx\left(r/\eta
\right)^2\left(1+6M/\eta \right)\ln A$. Due to the smallness of
$A$, the potential tends again, in the large $r$ limit, to
$-\infty $.

By fixing the parameter $B$ and increasing $A$ from small negative
values to zero, we deepen the potential well between the
marginally stable orbit (indicated by the peak in the left hand
side), and the marginally bound orbit $r_{mb}$ (at the local
maximum in the right hand side of the potential well). At $A=0$ we
reach the Schwarzschild geometry and for $A>0$ we obtain deeper
potential well, and reduce the local maximum further till it
becomes an inflexion point. As a result, for any non zero value of
$A$, we can only decrease the radial distance between $r_{ms}$ and
$r_{mb}$, as compared to the case of the Schwarzschild potential.
Therefore, the domain between the two radii for the Keplerian
motion of the matter will be decreased as well.

\begin{figure}[!ht] \centering
\includegraphics[width=.48\textwidth]{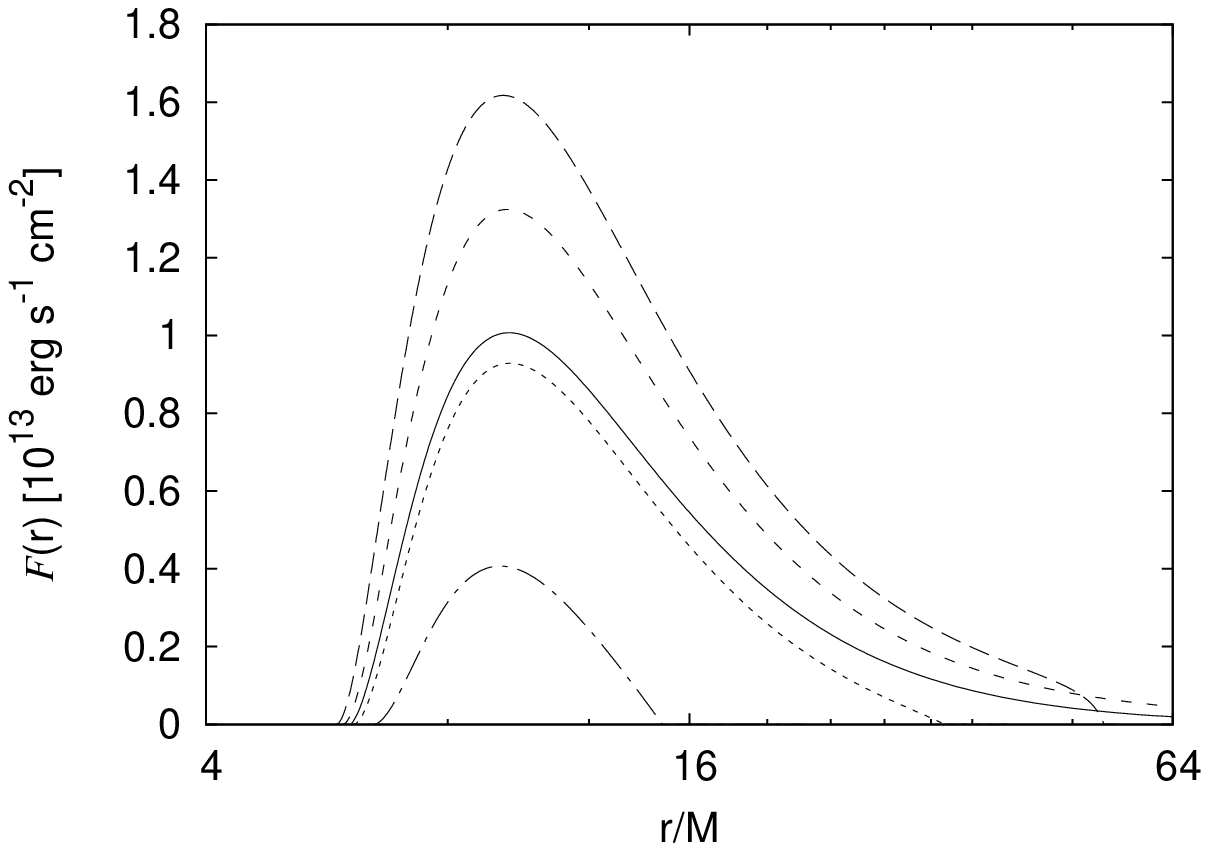}
\includegraphics[width=.48\textwidth]{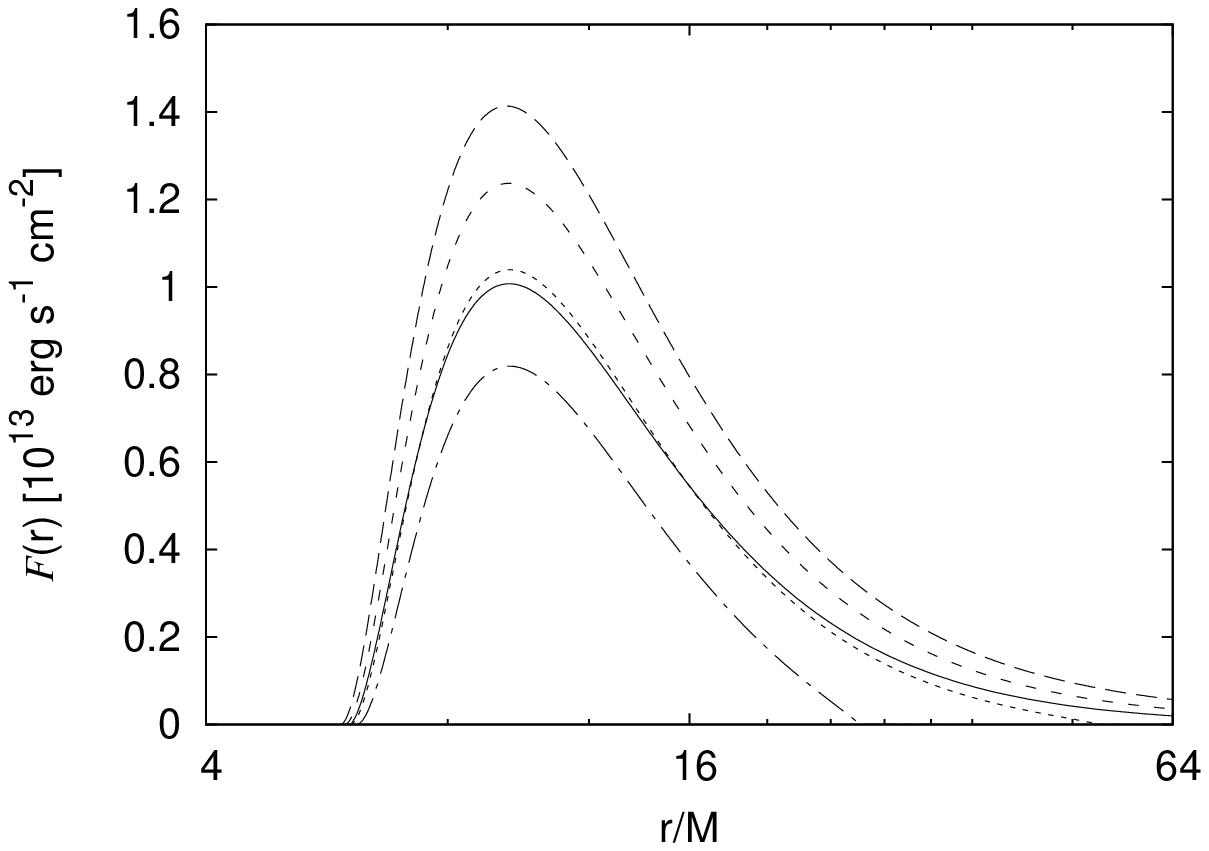}\\
\includegraphics[width=.48\textwidth]{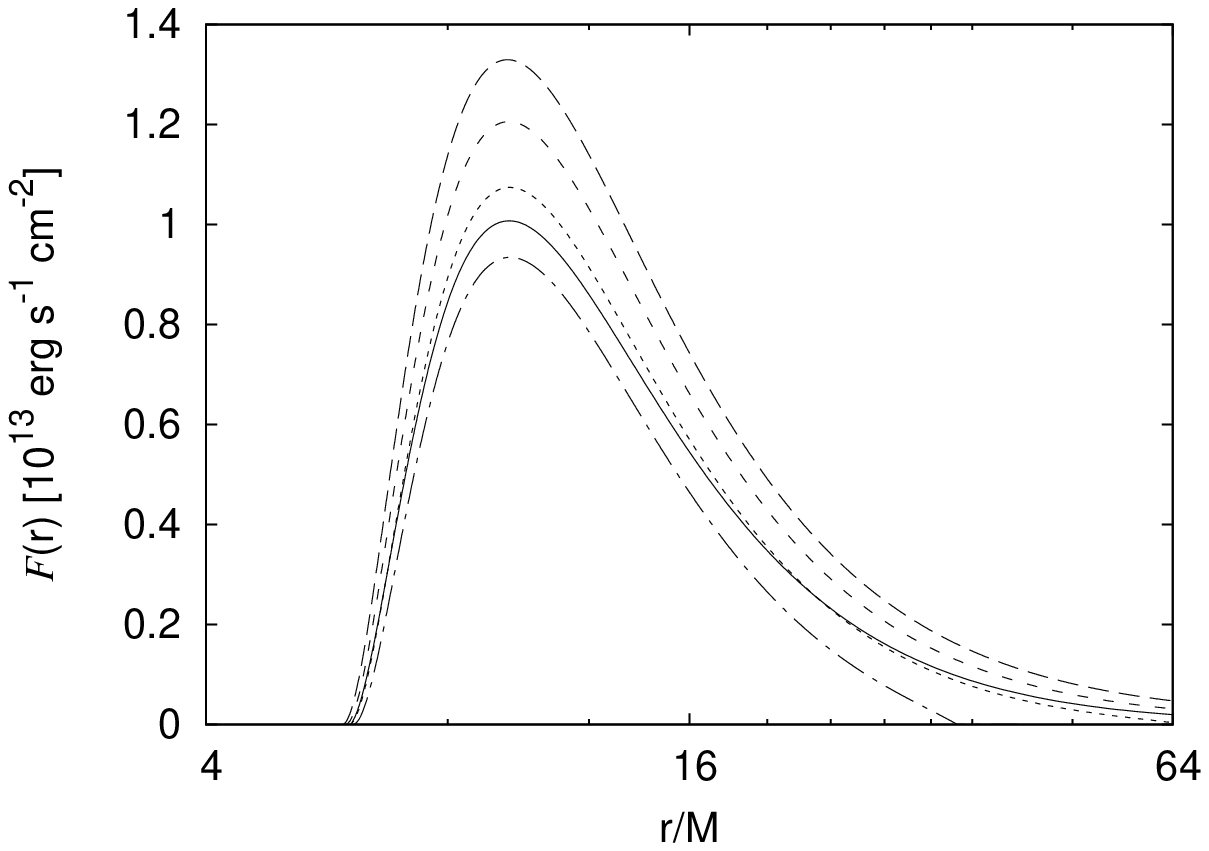}
\includegraphics[width=.48\textwidth]{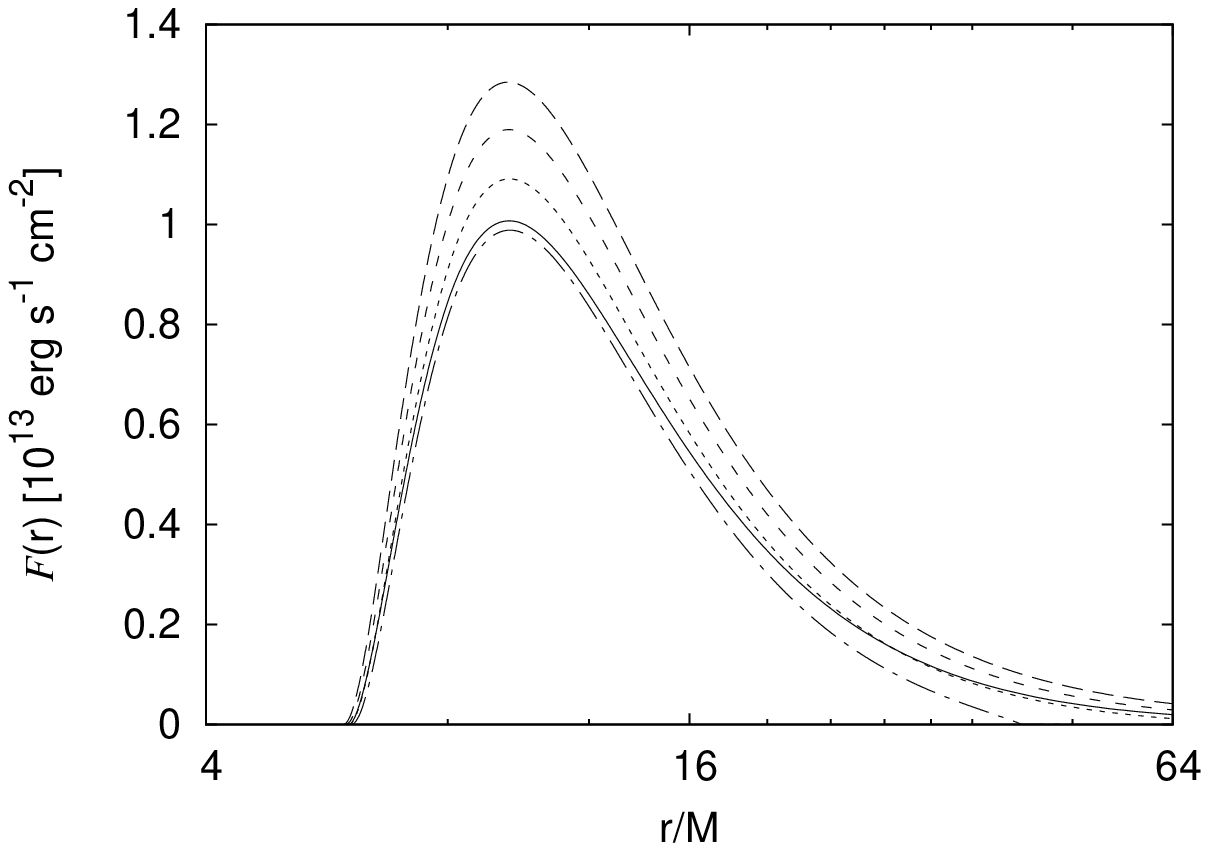}
\caption{The time averaged flux ${\mathcal F}(r)$ radiated by the
disk for a black hole of total mass $M=2.5\times10^6
M_{\bigodot}$. The parameter $B$ is set to $M$ (the upper left
hand plot), $2M$ (the upper right hand plot), $3M$ (the lower left
hand plot) and  $4M$ (the lower right hand plot), respectively.
The solid line is the energy flux for a Schwarzschild black hole
with total mass $M$. The various values of $A$ are
$-3\times10^{-3}$ (long dashed line), $-10^{-3}$ (short dashed
line), $10^{-3}$ (dotted line) and $3\times10^{-3}$ (dot-dashed
line).} \label{fR_F}
\end{figure}

The physical implications of this effect, specific to modified
$f(R)$ gravity models,  can be studied in detail in Figs.
\ref{fR_F}, where we have plotted the energy flux emitted by the
accretion disk for the same values of metric parameters $A$ and
$B$.

The cut-offs of the radial flux profiles in the right hand side at
$r_{mb}$ always appear, no matter if the values of $A$ are
positive or negative. However, the sign of the values of $A$ is
already important for the direction of the change of the
intensity, as compared to the Schwarzschild geometry. Since the
potential well is deeper for $A>0$, the specific energies of the
orbiting particles are lower, which decreases the radiated flux
over the disk surface. For negative values of $A$ we have the
opposite effect: in the higher potential well the particles have
higher specific energies, and more energy is radiated away than in
the previous case.

\begin{figure}[!ht]
\centering
\includegraphics[width=.48\textwidth]{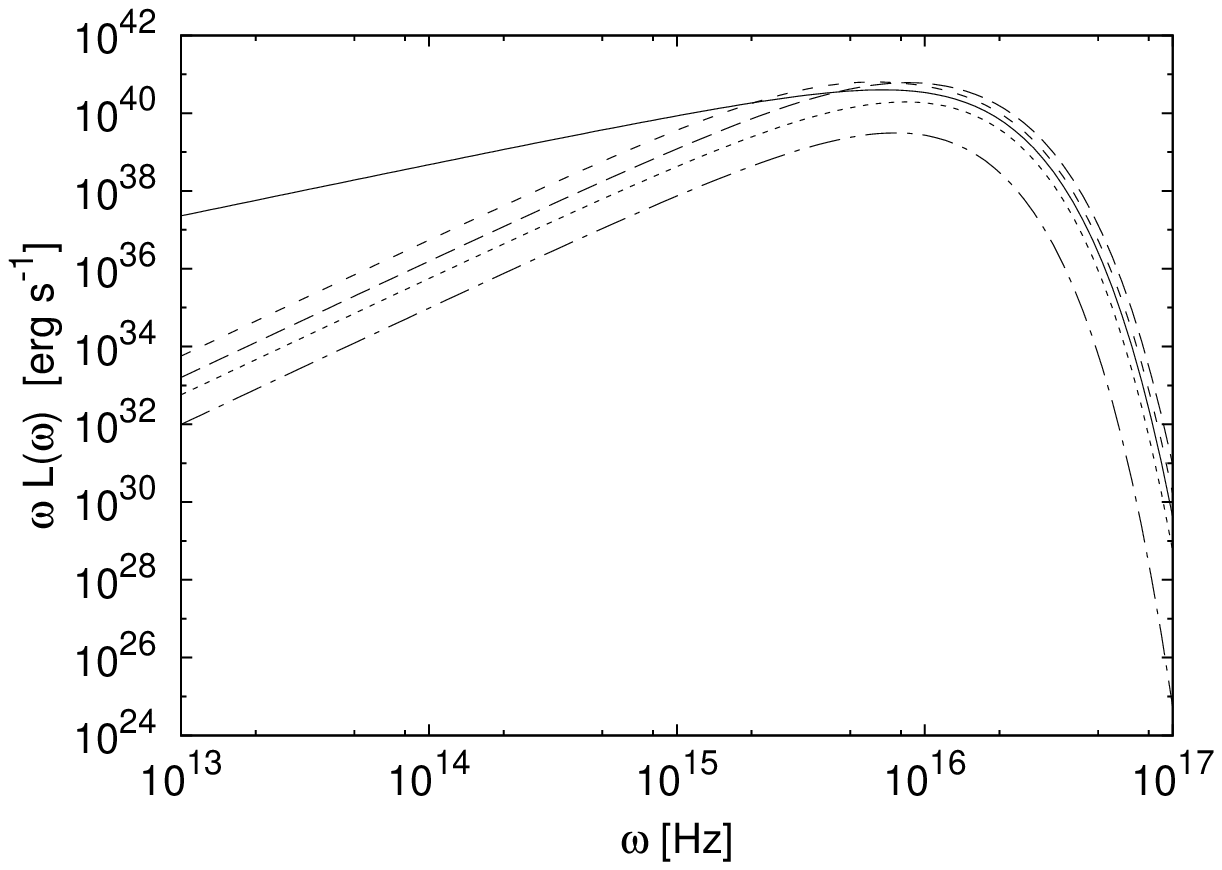}
\includegraphics[width=.48\textwidth]{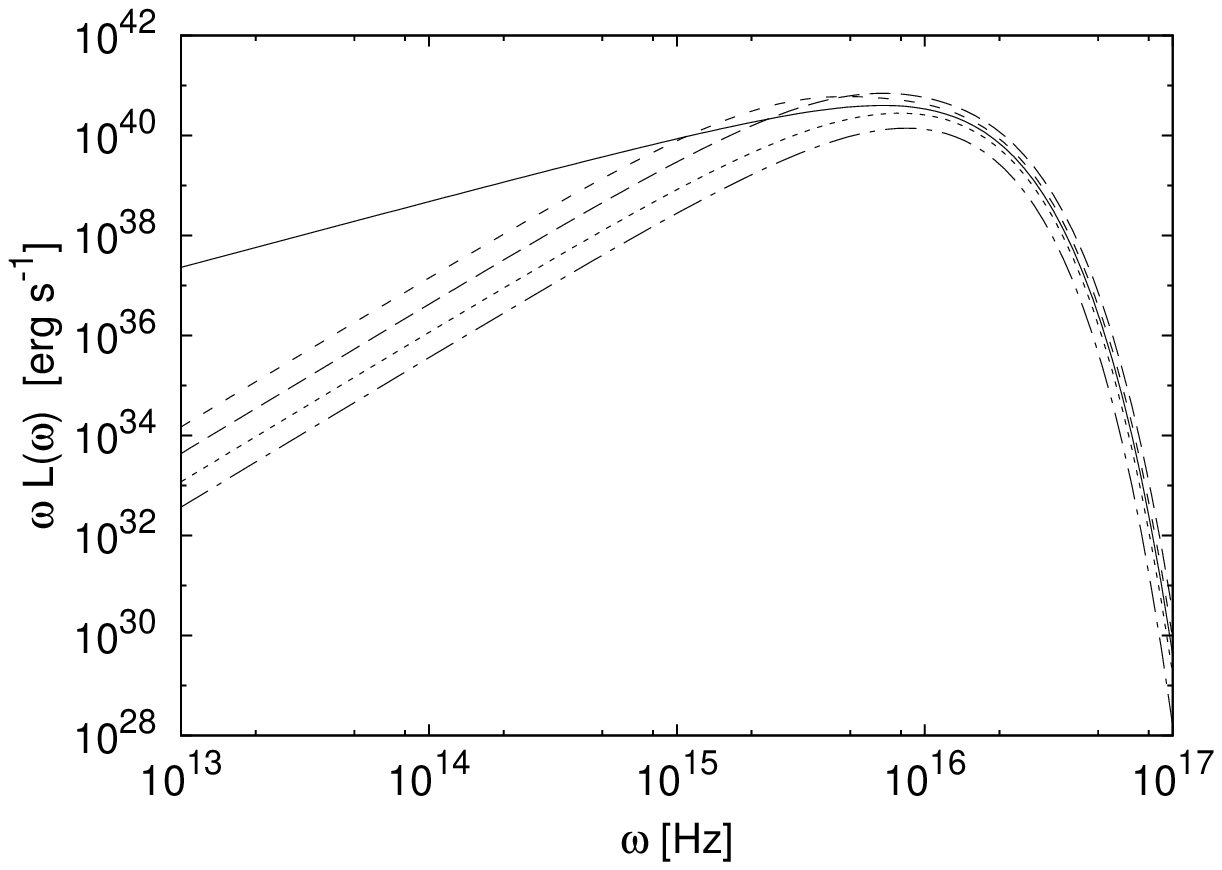}\\
\includegraphics[width=.48\textwidth]{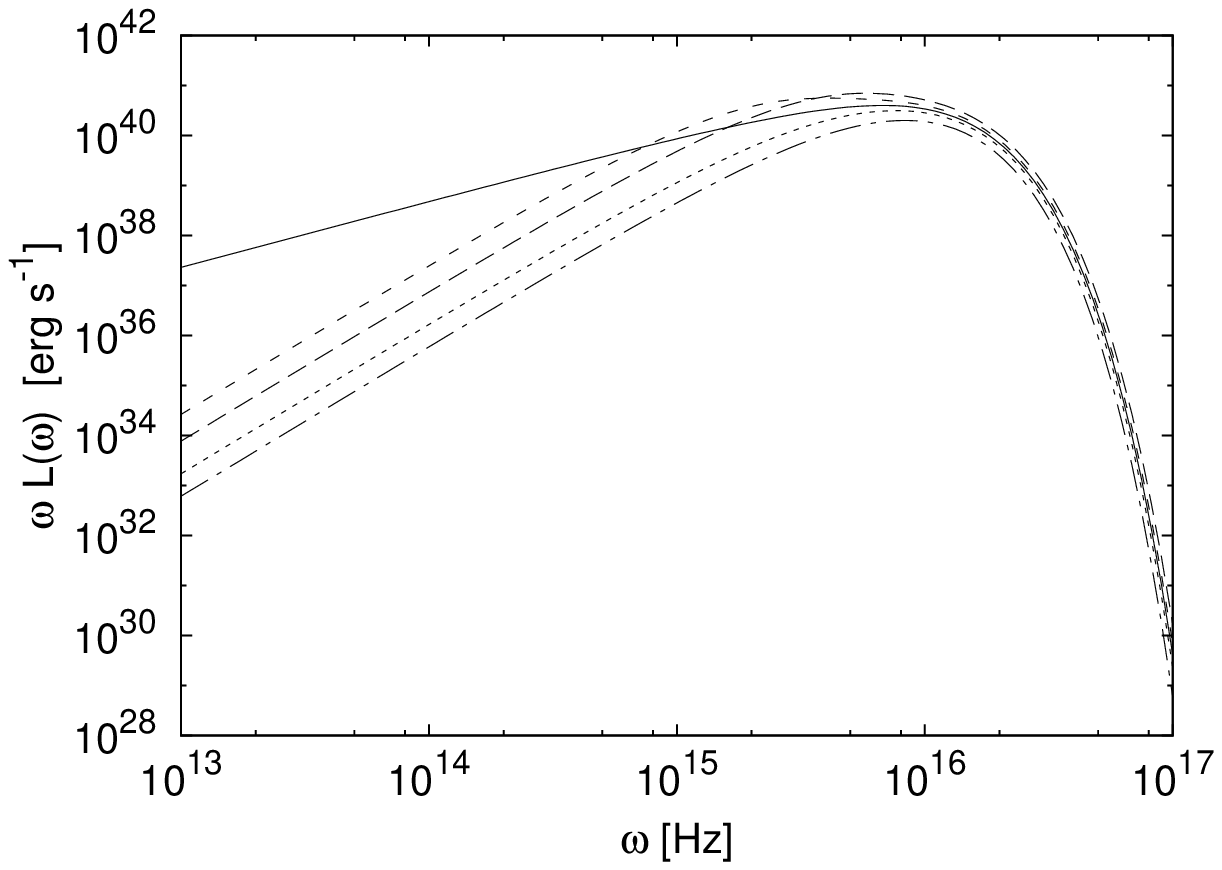}
\includegraphics[width=.48\textwidth]{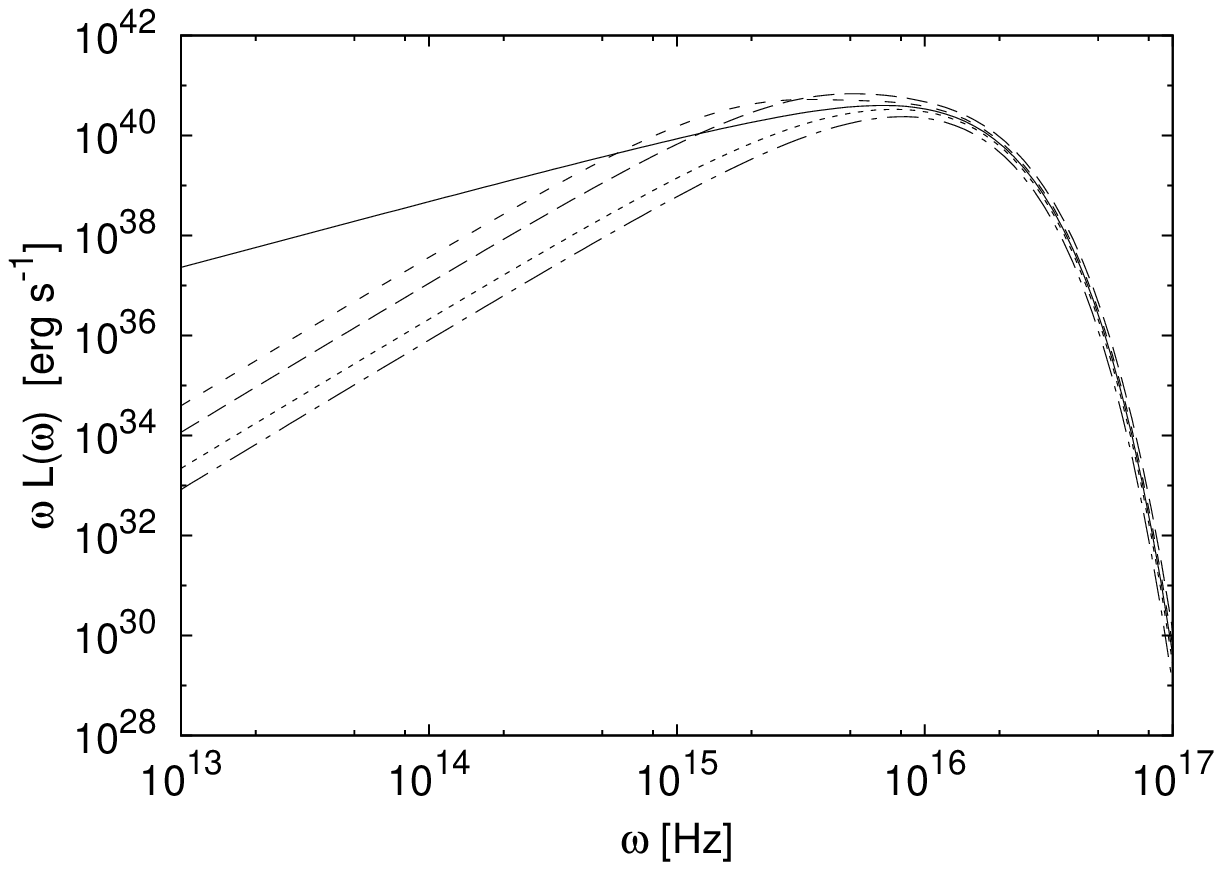}
\caption{Emission spectra $\omega L\left(\omega \right)$ of disks for a black hole of total mass $%
M=2.5\times10^6 M_{\bigodot}$. The accretion disk is aligned into
a face-on position, with $\cos i = 1$. The parameter $B$ is set to
$M$ (the left hand upper plot), $2M$ (the upper right upper plot),
$3M$ (the upper left hand plot) and $4M$ (the lower right hand
plot), respectively. The solid line is the emission spectrum for a
Schwarzschild black hole, with the same total mass $M$. The
various values of $A$ are $-3\times10^{-3}$ (long dashed
line), $-10^{-3}$ (short dashed line), $10^{-3}$ (dotted line) and $%
3\times10^{-3}$ (dot-dashed line).} \label{fR_L}
\end{figure}

These effects in the disk radiation can also be observed in the
emission spectrum of the accretion disk, plotted in Figs.
\ref{fR_L}.

In the lower regime, up to the cut-off frequency of the spectrum
at about $10^{16}$ Hz, the disk emission is much lower for $f(R)$
black holes than for the Schwarzschild black hole. As $A$ tends to
zero, the intensity is of course raising up, and approaches the
solid line representing the case of the Schwarzschild black hole.
For frequencies higher than the cut-off value, the differences are
not so striking, but still observable. For negative values of $A$,
the cut-off in the spectrum shifts to higher frequencies, whereas
positive values of $A$ produce lower cut-off frequency. The effect
of varying the parameter $B$ is also important, as all these
differences are more striking for lower values of $B$, as can be
seen by comparing the upper left hand plot, for $B=M$, with the
lower right hand one, corresponding to $B=4M$. For all the cases
we have considered here, the smaller values of the emission at
lower frequencies have the consequence of hardening the disk
emission spectrum. The emission spectrum is slightly modified due
to the effects of varying $A$, from small negative values to
positive ones, which softens the emission as we reach positive
values of $A$.

We also present the conversion efficiency $\epsilon$ of the
accreting mass into radiation, measured at infinity, which is
given by Eq. (\ref{epsilon}), for the case where the photon
capture by the hole is ignored. The value of $\epsilon$ measures
the efficiency of energy generating mechanism by mass accretion.
The amount of energy released by matter leaving the marginally
stable orbit, and falling down the black hole, is the binding
energy $\widetilde{E}_{ms}$ of the black hole potential. For
different metric parameters $A$ and $B$, the values of
$\widetilde{E}_{ms}$ are given, together with the radii of the
marginally stable orbits, in Table~\ref{Efficiency}, where the
figures corresponding to the Schwarzschild black hole appear in
the first line. For $A<0$ both the $r_{ms}$ and $\epsilon$ are
smaller than those for the Schwarzschild potential, whereas we
obtain higher values than $6$ and 0.0572 for $A>0$. Nevertheless,
the efficiency is decreasing with the increasing values of $B$.
For $B=4M$ the values of $\epsilon$ are about 10 percent smaller
than those for $B=M$.

\begin{table}[tbp]
\begin{center}
\begin{tabular}{|c|c|c|c|}
\hline
  $A$ & $B$ [$M$] & $r_{ms}$ [$M$] & $\epsilon$\\
\hline
0 & - & 6.0000 & 0.0572 \\
\hline
 $-3\times10^{-3}$ & 1 & 5.8154 & 0.0454 \\
\hline
 $-10^{-3}$ & 1 & 5.9218 & 0.0531 \\
\hline
 $10^{-3}$ & 1 & 6.1044 & 0.0616 \\
\hline
 $3\times10^{-3}$ & 1 & 6.4706 & 0.0712 \\
\hline
 $-3\times10^{-3}$ & 2 & 5.8861 & 0.0510 \\
\hline
 $-10^{-3}$ & 2 & 5.978 & 0.0551 \\
\hline
 $10^{-3}$ & 2 & 6.0489 & 0.0593 \\
\hline
 $3\times10^{-3}$ & 2 & 6.1700 & 0.0638 \\
\hline
 $-3\times10^{-3}$ & 3 & 5.9218 & 0.0530 \\
\hline
 $-10^{-3}$ & 3 & 5.9759 & 0.0558 \\
\hline
 $10^{-3}$ & 3 & 6.0398 & 0.0586 \\
\hline
 $3\times10^{-3}$ & 3 & 6.1044 & 0.0615 \\
\hline
 $-3\times10^{-3}$ & 4 & 5.9398 & 0.0540 \\
\hline
 $-10^{-3}$ & 4 & 5.9850 & 0.0561 \\
\hline
 $10^{-3}$ & 4 & 6.0306 & 0.0583 \\
\hline  $3\times10^{-3}$ & 4 & 6.0766 & 0.0604\\
\hline
\end{tabular}
\end{center}
\caption{The marginally stable orbit and the efficiency for
different $f(R)$ black hole geometries. The case $A=0$ corresponds
to the standard Schwarzschild general relativistic black hole.}
\label{Efficiency}
\end{table}

\section{Discussions and final remarks}

In the present paper we have considered the basic physical properties of
matter forming a thin accretion disc in the Schwarzschild type vacuum space-time metric of the $%
f(R)$ modified gravity models. The physical parameters of the disc
- effective potential, flux and emission spectrum profiles - have
been explicitly obtained for several values of the parameters
characterizing the vacuum solution of the generalized field
equations. All the astrophysical quantities, related to the
observable properties of the accretion disc, can be obtained from
the black hole metric. Due to the differences in the space-time
structure, the modified $f(R)$ gravity black holes present some
very important differences with respect to the disc properties as
compared to the standard general relativistic Schwarzschild case.

The determination of the accretion rate for an astrophysical
object can give a strong evidence for the existence of a surface
of the object. A model in which Sgr A*, the $3.7\times
10^{6}M_{\odot }$ super massive black hole candidate at the
Galactic center, may be a compact object with a thermally emitting
surface was considered in \cite{BrNa06}. For very compact surfaces
within the photon orbit, the thermal assumption is likely to be a
good approximation because of the large number of rays that are
strongly gravitationally lensed back onto the surface. Given the
very low quiescent luminosity of Sgr A* in the near-infrared, the
existence of a hard surface, even in the limit in which the radius
approaches the horizon, places a
severe constraint on the steady mass accretion rate onto the source, ${\dot{M%
}}\leq 10^{-12}M_{\odot }$ yr$^{-1}$. This limit is well below the
minimum accretion rate needed to power the observed submillimeter
luminosity of Sgr A*, ${\dot{M}}\geq 10^{-10}M_{\odot }$
yr$^{-1}$.

Thus, from the determination of the accretion rate it follows that
Sgr A* does not have a surface, that is, it must have an event
horizon. Therefore the study of the accretion processes by compact
objects is a powerful indicator of their physical nature. Since,
as one can see from Table~\ref{Efficiency}, the conversion
efficiency in the case of the $f(R)$ vacuum solutions is different
as compared to the general relativistic case, the determination of
this parameter could discriminate, at least in principle, between
the different gravity theories, and constrain the parameters of
the model.

\section*{Acknowledgments}

The work of T. H. is supported by an RGC grant of the government
of the Hong Kong SAR. Z. K. is indebted to the colleagues in the
Department of Physics of the University of Hong Kong for their
support and warm hospitality during the preparation of this work.

\end{document}